\documentclass[twocolumn,showpacs,preprintnumbers,amsmath,amssymb]{revtex4}
\usepackage{dcolumn}% Align table columns on decimal point
\usepackage{bm}% bold math
\usepackage[dvipdf]{graphicx}% Include figure files
\DeclareGraphicsExtensions{.jpg,.pdf,.mps,.png,.eps,.ps,.EPS}
               
\begin{document}
\def\be{\begin{equation}}
\def\ee{\end{equation}}
\def\bc{\begin{center}}
\def\ec{\end{center}}
\def\bea{\begin{eqnarray}}
\def\eea{\end{eqnarray}}
\newcommand{\hT}{ {\cal T} }
\newcommand{\hB}[2]{ {\cal B} \left[ {#1} \cdot {#2} \right]}
\newcommand{\avg}[1]{\left\langle {#1} \right\rangle}

%\draft \twocolumn[\hsize\textwidth\columnwidth\hsize\csname
%@twocolumnfalse\endcsname
\title{Self-organized critical network dynamics}
\author{Ginestra Bianconi and Matteo Marsili}
\affiliation{The Abdus Salam International Center for Theoretical Physics, 
Strada Costiera 14, 34014 Trieste, Italy 
}
 
\begin{abstract}
We propose a simple model that aims at describing, in a stylized
manner, how local breakdowns due unbalances or congestion propagate in
real dynamical networks. The model converges to a self-organized
critical stationary state in which the network shapes itself as a
consequence of avalanches of rewiring processes. Depending on the
model's specification, we obtain either single scale or scale-free
networks. We characterize in detail the relation between the
statistical properties of the network and the nature of the critical
state, by computing the critical exponents. The model also displays
a non-trivial, sudden, collapse to a complete network.
\end{abstract}
\maketitle

Complex networks underlying many social and technological systems is a
subject of booming recent interest \cite{AB,Doro_rev}. On one side the
structure of such networks has non-trivial properties \cite{AB}, which
influences dramatically the nature of processes taking place on them
(see e.g. \cite{Optimal,Epidemics}). On the other, the network's
structure constrains in a peculiar way the growth \cite{ABGrowth} and
evolution \cite{Ebel,Social} of the network itself.  This calls for an
extension of statistical physics, which traditionally studies
collections of dynamical variables interacting through a fixed
network, to systems where the network of interactions itself becomes a
dynamical variable.

Here we focus on dynamical networks where links do not represent
physical bonds but rather relationships or communication channels.
Ref. \cite{Optimal} suggests that a structured network of
communications aimed at solving problems or carrying out specific
functions is a crucial feature of firms and organizations in
general. Routers table in the Internet is also an example of a
dynamic communication network. The network of financial institutions,
linked by mutual contracts and loans, provides a further example of
a dynamic network. Beyond ``static'' design problems, such
as e.g. minimizing congestion \cite{Optimal} or redistributing
optimally the loads \cite{Moreno}, systems of this type also pose
``dynamical'' problems such as how and to what extent does congestion
or breakdown events propagate through the system. 

Here we address these problems in a dynamic network subject to two
competing forces: on one side, there is a drive toward increasing
complexity, by e.g. adding new links, because the system performs more
efficiently its functions as it becomes more densely
interconnected. On the other, the resulting increase in complexity may
bring about conflicting constraints, imbalances or congestion problems
which may cause a local breakdown of the network. A local breakdown
may engender a re-adaptation in its neighborhood which may
inadvertently cause the breakdown to propagate further on the network.

For example, a change in some router's table in the Internet, which is
meant to improve efficiency, may inadvertently cause congestion at
some node downstream. This may trigger several other changes in that
local neighborhood, as routers try to avoid the congested nodes. But
these changes may, in their turn, cause further congestion elsewhere,
and the problem may expand even further, as an avalanche, to a wider
region. The stipulation of a contract between two institutions, which
is in principle beneficial to both, may also increases their operative
constraints, making them less adaptable to a changing environment and
hence more exposed to the risk of bankruptcy. The failure of one
institution, likely induces a rearrangement of the institutions linked
with it and perhaps engenders effects which propagate further
across the network \cite{Holyst}. Similar phenomena may take place 
in social or trade networks.

Rather than trying to model in a realistic manner one of the problems
just discussed, we focus on a simple model of network dynamics which
captures the two main ingredients discussed above: a slow dynamics
where links are added to the network and a fast relaxation dynamics of
avalanche events. The motivation for this choice, is that, the
detailed understanding of the behavior of a simple model with these
features, may be the basis or at least a guide for addressing more
complex and realistic situations, such as those discussed above.  Our
main finding is that such systems can self-organize close to a
critical point where each modification of the network's architecture
can have unforeseable consequences which possibly affect a wide region
of the system. This may have some bearing on the intermittence of
Internet traffic or on the nature of financial crises and
recessions. 

We start from a empty network of $N$ nodes in which every node $i$ is
assigned a fitness $f_i$ drawn from a probability distribution
$\rho(f)$. Let $F_{i}$ be the set of neighbors of $i$ and $k_i=|F_i|$
be the number of neighbors of $i$. At every time step a link is added
between two previously unconnected random nodes $i$ and $j\not\in
F_i$. With probability $f_j$ nothing happens whereas with probability
$1-f_j$ the node $j$ becomes unstable or congested and it
``topples''. As a result all its links (including that with $i$) are
rewired to randomly chosen nodes, i.e. for any $h\in F_j$, a node
$l\not\in F_h$ is chosen at random and the link $jh$ is rewired to
$hl$. In its turn, with probability $1-f_l$, also node $l$ may become
unstable and topple. Hence toppling of node $j$ may start an
avalanches of toppling events which propagates through the network
rearranging it. Unstable nodes, after they topple, remain unconnected
from the network \cite{notacon} and are assigned a new fitness value
drawn from the distribution $\rho(f)$. Hence toppling is equivalent to
replacing the unstable node with a new one.

In order to stabilize the network and reach a stationary state, we
introduce dissipation of links: at each toppling event, with
probability $\lambda$ all the links of the unstable node are removed
from the network. Note that without dissipation the number of links
would increase in time until the complete graph is reached. The
complete graph ($j\in F_i$ $\forall i,j,~i\neq j$) is an absorbing
state of the dynamics, because no link can be added to it.

The distribution $\rho(f)$ is the only parameter of the model. An
alternative class of models can be defined by specifying the
probability $u_k$ that a node with $k$ neighbors becomes unstable upon
addition of a further link. A relation between the two models is
possible, along the lines of Ref. \cite{Polya}, in the limit
$\lambda\to 0$ on no dissipation. Then the probability to find a node
with $k$ neighbors and $f_i\in [f,f+df)$ is
$\rho(f|k)df\propto f^k\rho(f)df$ where $f^k$ is the probability that a 
node with $f_i=f$ has $k$ neighbors. Then a model with

\be
u_k=\int_0^1\!df(1-f)\rho(f|k)=\frac{\int_0^1\!df(1-f)f^k\rho(f)}{\int_0^1
 df f^k\rho(f)} 
\label{uk}
\ee

\noindent
is completely equivalent to one specified in terms of $\rho(f)$, in
the limit $\lambda\to 0$.
The dependence on $k$ of $u_k$ reflects the fact that $k_i$ and $f_i$
are positively correlated because nodes with higher fitness have
smaller chance of becoming unstable. For convenience we shall refer
mostly to models specified in terms of $u_k$ using Eq. (\ref{uk}) to
translate the results in the original model.

\begin{figure}
\includegraphics[width = 52 mm, height = 33 mm ]{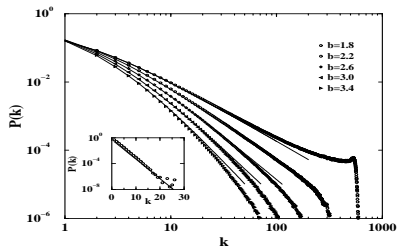} 
\caption{Degree distribution $p_k$ for network of $N=10^4$ nodes
  (averaged over $300 N$ time steps after equilibration) for the model
  of Eq. (\ref{pow}) and $b=1.8,\ldots,3.4$ and $\lambda=5\cdot
  10^{-3}$ and for $u_k=u=0.5$ (inset). The solid lines corresponds to
  the theoretically expected $p_k$ in the limit
  $\lambda\rightarrow 0$ [Eq.~$(\ref{p_k.eq})$ and $p_k=u(1-u)^k$
  respectively].}  
\label{pk.fig}
\end{figure}

\begin{figure}
\includegraphics[width = 52 mm, height = 33 mm ]{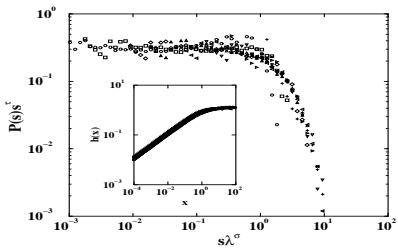} 
\caption{Data collapse of the avalanche distribution $P(s)$ with $s>7$
  for networks of $N=10^3$ nodes averaged over $200N$ time steps after
  equilibration. Here $b=2.5$ and
  $\lambda=0.01,0.02,0.04,0.06,0.18,0.31$. The data collapse was done
  taking the theoretical values $\tau=3/2$ and $\sigma=2$ of the
  exponents. Inset: the scaling function $h(x)$ for the same data
  set. }
\label{ps.fig}
\end{figure}

Our results can be summarized as follows: {\em i)} 
when $u_k$ decays faster than $1/k$ there is a
critical $\lambda_c$ such that the network evolves toward a complete
graph for $\lambda<\lambda_c$. The same happens for $u_k\simeq b/k$
($k\gg 1$) and $b<3/2$ whereas when $b>3/2$ or when $u_k$ decays
slower than $1/k$, the collapse takes place only in finite
networks. Indeed we find complete graphs only for
$\lambda<\lambda_c\sim N^{-\gamma}$, where $\gamma=\frac{2b-3}{b-1}$
for $3/2<b<2$ and $\gamma=1$ otherwise. {\em ii)} the non-collapsed phase
$\lambda>\lambda_c$ is characterized by an uncorrelated random network
\cite{NewmanRand} with finite average degree and a degree 
distribution $p_k$ which depends on $\rho(f)$ (or $u_k$) (see
Fig. \ref{pk.fig}). In particular, if $u_k$ decays slower than $1/k$
then $p_k$ decays faster than any power, whereas if $u_k\simeq b/k$
then $p_k\sim k^{-b}$. {\em iii)} The 
dynamics converges to a stationary sequence of avalanches of rewiring
processes with a power law distribution $P(s)\sim s^{-\tau}$ of sizes (see
Fig. \ref{ps.fig}). As in
Ref. \cite{Polya}, we shall define the size $s$ of an avalanche as the
number of toppling events that it causes.
The exponent $\tau$ takes the mean field value\cite{MF}
$\tau=3/2$ when $u_k$ decays slower than $2/k$, whereas $\tau=1+1/b$
if $u_k\simeq b/k$ with $3/2<b<2$.

The collapse to a complete graph, where congestion is minimal, is
reasonable, given that the network is trying to adapt by avoiding
congestion. The non-trivial issue on which we shall concentrate mostly, 
is related to the self-organized critical state.
In view of the special role played by the case $u_k\sim 1/k$, we focus
on the following simple forms of $\rho(f)$ and to the corresponding
$u_k$: 

\bea 
\rho(f)=b (1-f)^{b-1},~ & ~u_k=\displaystyle{\frac{b}{b+k+1}}
\label{pow}
\eea

In order to shed light on the model's behavior, let us derive an equation for
the avalanche distribution. Define $s_k$ as the avalanche size
originating from an unstable node with $k$ neighbors. This is a random
variable which can be decomposed as follows
\be
s_k=1+d\sum_{j=1}^{k+1} v_{k_j} s_{k_j}.
\label{avala}
\ee

\noindent
into the contribution of the unstable node and those of the
avalanches $s_{k_j}$ ensuing from its neighbors, with $k_j$ being the
number of neighbors of the $j^{\rm th}$ neighbor. Note that the sum
runs over $k+1$ links as it includes the link which caused the
instability and the $k$ pre-existing neighbors. In Eq. (\ref{avala})
$d=0,1$ describes the effect of dissipation with
$P(d=0)=\lambda=1-P(d=1)$ whereas $v_{k_j}$ takes value $v_{k_j}=0$ if
rewiring of the link to the $j^{\rm th}$ neighbor causes no further
toppling, and $v_{k_j}=1$ otherwise. Hence $P(v_l=1) =1-P(v_l=0)=
u_l$.

Now we can write the generating function $\phi_k(z)=\avg{z^{s_k}}$ of
the probability $P(s|k)$ to have an avalanche of size $s$ given that
the initiator node has $k$ neighbors. From this, it is easy to find
the generating function $\chi(z)$ of the distribution $P(s)$ of
avalanche sizes 
\be
\chi(z)\equiv \sum_{s} P(s) z^s= 
\frac{1}{\bar u}\sum_{k=0}^\infty p_k u_k \phi_k(z).
\label{defchi}
\ee
with $\bar u=\sum_k p_k u_k$. After some algebra, using the fact that
the rewired nodes are chosen randomly in the network, Eq. (\ref{avala})
leads us to 
\be
\chi(z)=z\left[
\lambda+\frac{1-\lambda}{\bar u}\sum_{k=0}^\infty p_k u_k\left[1-\bar
  u+\bar u\chi(z)\right]^{k+1}\right]
\label{chi.eq}
\ee
which is a non-linear self-consistent equation for $\chi(z)$.
Note that $\chi(1)=1$ as it should and that
\be
\chi'(1)=\avg{s}=\frac{1}{1-(1-\lambda)\avg{(k+1)u}}
\label{s_mean.eq}
\ee

Apart from the parameters $\lambda$ and $u_k$, Eq. (\ref{chi.eq})
depends also on the stationary state degree distribution $p_k$. Hence,
before discussing Eq. (\ref{chi.eq}) further, we need to elaborate on
the nature of the stationary state.  A necessary condition in order to
be at the stationary state is that the total number of links $K(t)$
present in the network is constant on average. $K(t)$ increases by one
for the random addition of a new link and is reduced by the total
number of links $\Lambda$ dissipated during the avalanche
that follows. Here 
\be 
\Lambda=v \sum_{i=1}^{s} d_i(k_i+1) 
\ee 
where $v=1$ if the chosen site is unstable and $v=0$ otherwise, and 
$d_i=1$ if dissipation occurs at the toppling site $i$, otherwise
$d_i=0$. Consequently $v$
and $d_i$ have average values $\bar{u}$ and $\lambda$ and thus 
$\avg{\Lambda}=\lambda \avg{(k+1)u_k}\avg{s}$. Then, stationarity
$\avg{\Lambda}=1$ and Eq. (\ref{s_mean.eq}) imply that
\be
\avg{(k+1)u_k}=1.
\label{eq.eq}
\ee  
As a byproduct, we find $\avg{s}=1/\lambda$, in perfect agreement with
numerical simulations.

The stationary degree distribution $p_k$ can be found by quantifying the
Markov chain of possible transitions $k_i\to k_i'$ during the
dynamics. In the limit $\lambda\to 0$ and
$N\to\infty$ where we can neglect dissipation and finite size
effects, there are only two processes which take place
on each node: $k_i\to k_i+1$, with probability $1-u_{k_i}$ and $k_i\to
0$ with probability $u_{k_i}$. Then, in the stationary state, $p_k$
satisfies 
\begin{equation}
p_{k+1}=(1-u_k)p_k.
\label{stat}
\end{equation}
Taking the sum of Eq. (\ref{stat}) on $k$ we find $p_0=\avg{u_k}$,
which means that the fraction of sites with no neighbor is equal to
the probability that a node becomes unstable. Furthermore multiplying
Eq. (\ref{stat}) by $k+1$ and taking the sum over $k$, we recover the
stationary condition (\ref{eq.eq}). In the simplest case  
$u_k=u$ for all $k$, we find $p_k=u(1-u)^k$ whereas with
Eq. (\ref{pow}) we find 
\be
p_k=(b-1)\frac{\Gamma(b)\Gamma(k+1)}{\Gamma(b+k+1)}\sim k^{-b}
\label{p_k.eq}
\ee 
where the asymptotic power law behavior holds for $k\gg
1$. Notice
that $\avg{k}=1/(b-2)$ diverges when $b\to 2^-$ and that there is a
finite fraction $\bar u=1-1/b$ of unconnected nodes. Still the network
has a giant connected component for $b<7$ \cite{NewmanRand}. Eq. (\ref{stat})
yields a $p_k$ which decays faster than any power if $u_k$ decays less
slowly than $1/k$, or if it increases. Conversely, if $u_k$ decays
faster than $1/k$, we find that $p_k$ is not normalizable for
$N\to\infty$. Numerical simulations (see Fig. \ref{pk.fig}) 
fully supports this picture, even though the
effects of dissipation and finite size are clearly evident for $k\gg
1$. 

The neglect of dissipation, when $N$ is finite, is a reasonable
approximation if nodes with maximal degree $k_i=N-1$ are not stable. 
A node connected to all neighbors, cannot
receive further links and hence cannot become unstable. Its degree
decreases only if dissipation occurs at a node connected to them. The 
rate of this process, for a node with degree $k$, is $\lambda
k/\avg{k}$. Hence if $\lambda N/\avg{k}\gg 1$, nodes with $k\sim N$
decay very fast and the only effect of dissipation is to introduce a
cutoff $k_c\sim 1/\lambda$ in the distribution $p_k$. When $\lambda
N/\avg{k}\sim 1$ we expect a transition (close) to the complete graph
$k_i=N-1$ for all $i$. 
When $\avg{k}$ is finite, i.e. for $b>2$ or for $u_k$ which decays
slower than $1/k$, the collapse to a complete graph takes place for
$\lambda<\lambda_c\sim N^{-1}$. When $3/2<b\le 2$ the average degree
$\avg{k}\sim N^{\frac{2-b}{b-1}}$ diverges with the system size. Then
the decay rate of totally connected nodes is $\sim\lambda N^{\gamma}$
with $\gamma=\frac{2b-3}{b-1}$ and the collapse to a complete graph
takes place for $\lambda<\lambda_c\sim N^{-\gamma}$. In
both cases (i.e for $b>3/2$) for any $\lambda<1$, it is always
possible to take $N$ large enough to make the decay rate $\lambda
N^{\gamma}$ large enough so that finite size effects can be
neglected. But when $b<3/2$ this is no longer true because
$\avg{k}\sim N$ and the decay rate of completely connected nodes
remains finite ($\gamma=0$) even when $N\to\infty$. Beyond a finite
dissipation rate $\lambda_c$, the network collapses to the complete
graph.
\begin{figure}
\includegraphics[width = 52 mm, height = 33 mm ]{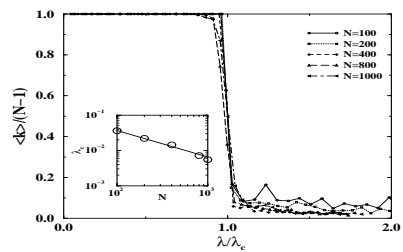} 
\caption{Average degree as a function of $\lambda$ for different
system sizes and $b=1.8$. The value of $\lambda_c$ at which the
transition takes place is plotted in the inset against $N$. The full
line is the theoretical prediction $\lambda_c\propto
N^{-\gamma}$ with $\gamma=0.75$ for $b=1.8$.}
\label{trans.fig}
\end{figure}
Fig. \ref{trans.fig} fully confirms the theoretical insight discussed
above. When $\lambda>\lambda_c$, where $\lambda_c\sim N^{-\gamma}$
(see inset), the dynamics reaches a network with finite average
degree, whereas for $\lambda<\lambda_c$ a collapse to the complete
graph is observed, with a transition which is sudden and
discontinuous.  In the case in which the $u_k=u$ we also observe a
transition from a finite average connectivity to a average
connectivity of order $N$. The transition occurs for values of
$\lambda_c\sim N^{-1}$ but the transition is rather smooth.
Having discussed the stationary state, let us go back to the
Eq. (\ref{chi.eq}) for the avalanche size distribution. We focus on
the region $1\gg \lambda\gg N^{-\gamma}$ where the network is not
densely connected and dissipation effects are weak.
Anticipating that the avalanche size distribution acquires a cutoff $s_c\sim
\lambda^{-\sigma}$, for some exponent $\sigma>0$, we postulate
the scaling form $P(s)\simeq s^{-\tau}\Phi(s\lambda^{\sigma})$ with 
finite $\Phi(0)$ and $\Phi(x)\to 0$ faster than any power as
$x\to\infty$. Such a scaling hypothesis is fully corroborated by
numerical results, as shown in Fig. \ref{ps.fig}.
This corresponds\cite{BMF} to an analogous scaling form 
\be
\chi(z)\simeq 1-(1-z)^{\tau-1}h\left(\frac{1-z}{\lambda^\sigma}\right)
\label{hx}
\ee
\noindent
for the generating function for $\lambda\ll 1$ and $1-z\sim
\lambda^\sigma$.  Setting for convenience $1-z=x\lambda^\sigma$,
asymptotic analysis for $\lambda\ll 1$ shows that the leading orders
of Eq. (\ref{chi.eq}) are
\be
\lambda^{1+\sigma(\tau-1)}x^{\tau-1}h(x)-\lambda^\sigma x+
c\lambda^{\sigma\beta(\tau-1)}x^{\beta(\tau-1)}h^\beta(x) =0
\label{scale}
\ee
where 
\bea
\beta=2,& c=\frac{b-1}{b(b-2)} & \hbox{~~for $b>2$}\\
\beta=b,& c=\frac{\pi b^{b-1}(1-b)^b}{\sin[\pi(2-b)]} & \hbox{~~for
  $b<2$}.
\eea
Note that $c\sim 1/|b-2|$ diverges when $b\to 2$ is approached from
both sides.
Dividing Eq. (\ref{scale}) by $\lambda^\sigma$ and taking the scaling
limit $\lambda\to 0$ with $x$ finite, we
find a non-trivial result with all three terms finite if we choose \begin{eqnarray}
\tau=3/2,& \sigma=2 & \hbox{for $b>2$}\\
\tau=1+1/b,& \sigma=b/(b-1) & \hbox{for $b<2$}.
\end{eqnarray}and the scaling function is the inverse of $x(h)=h^\beta/[1-ch^\beta]^\beta$. In particular $h(x)\to c^{1/\beta}$ for $x\to\infty$ and $h(x)\sim x^{1/\beta}$ for $x\ll 1$. For $b>2$
the solution coincides with that of other mean field models \cite{BMF}
$h(x)=[\sqrt{{c}/{x}+4}-\sqrt{{c}/{x}}]/(2c)$
and perfectly matches numerical simulations for a range of values of
$\lambda$ (see the inset of Fig. \ref{ps.fig}). It is easy to check
that the model with $u_k$ decaying slower than $1/k$ falls in the
$\beta=2,~\tau=3/2$~$\sigma=2$ universality class.
In conclusion we have shown how a slow growth dynamics and a fast
relaxation through avalanche events can generate a dynamical 
network with given degree distribution. The stationary state
is critical in the sense that the avalanches of all sizes occur, and 
is reached spontaneously without fine tuning of external parameters
as long as the dissipation rate is larger than a given threshold
$\lambda_c\sim N^{-\gamma}$. For smaller dissipation rates, the
network collapses to the complete graph. While the detailed solution
depends on the particular simplicity of the model chosen, the generic
picture may apply to a wider class of systems and capture some
features of the non-linear and intermittent behavior of real systems. 
We acknowledge F. Vega-Redondo for useful comments and  discussions.
This work is supported in part by the European Community's
Human Potential Programme under contract HPRN-CT-2002-00319,
STIPCO.


\begin{thebibliography}{}
\bibitem{notacon}
If the neighbor $k$ of an unstable node $j$ is connected to
all the nodes, the link $jk$ can only be rewired to node $j$
again.
\end{thebibliography}
\end{document}